\documentclass[12pt]{article}
\usepackage[english]{babel}
\usepackage{amssymb}
\def\vE{\vec{E}}
\def\VE*{\vec{E}^{*}}

\def\vA{\vec{A}}
\def\vx{\vec{x}}

\def\vk{\vec{k}}
\def\bfa{{\bf a}}
\def\nn{\nonumber}
\def\ba{\begin{eqnarray}}
\def\ea{\end{eqnarray}}
\begin{document}
\title{{\bf Huyghens Principle, Planck Law:
Peculiarities in the Behavior of Planar Photons\\}
\author{Winder A. Moura-Melo$^{\mbox{a,}}$ \thanks{Email:
winder@stout.ufla.br} \hspace{.1cm}and J.A.
Helay\"el-Neto$^{\mbox{b,c,}}$\thanks{Email: helayel@cbpf.br,
helayel@gft.ucp.br}
 \\ \\
$^{\mbox{a}}$\hspace{.1cm}Departamento de Ci\^encias Exatas,
Universidade
Federal de Lavras\\Caixa Postal 37, Cep 37200-000, Lavras, MG,
Brasil.\\
\\$^{\mbox{b}}$Centro Brasileiro de Pesquisas F\'{\i}sicas
\\ Rua
Xavier Sigaud 150 - Urca, Cep 22290-180, Rio de Janeiro, RJ, Brasil.\\
\\$^{\mbox{c}}$Grupo de F\'{\i}sica Te\'orica,
Universidade
Cat\'olica de Petr\'opolis\\Av. Bar\~ao do Amazonas 124, Cep
25685-070, Petr\'opolis, RJ,
Brasil.}}
\date{}
\maketitle
\begin{abstract}
Huyghens principle and Planck law
are studied in Maxwell and
Maxwell-Chern-Simons frameworks in (2+1)
dimensions. Contrary to (3+1) dimensions, massless
photons are shown to violate Huyghens principle in planar world. In
addition, we obtain that Planck law is no longer proportional to
$\nu^3$, but to the squared frequency, $\nu^2$, of the planar photons.
We also briefly discuss possible physical consequences of these
results.  \end{abstract}\newpage

\section{Introduction}
Field theories lying in three space-time dimensions have attracted a
great deal
of efforts since nearly two decades \cite{DJTetc}. Such studies have
given us a
number of new theoretical insights which have been useful for a better
theoretical understanding of several aspects, not only related to these
particular theories, but also to a wider class of models, including
some ones
defined in other dimensions. On the other hand, planar physics has also
been
shown to provide good explanations of important phenomena observed in
the
Condensed Matter domain, among them, those concerned to the Quantum
Hall
Effect\cite{qhe} and to the High-Tc Superconductivity
\cite{htcsup,anyonsup}.\\

A quite important underlying aspect of these theoretical
applicabilities is
undoubtely a number of peculiar characteristics exhibited by such models,
among them we may quote charge and spin fractionization
\cite{anyonsup,frac}. Actually, these interesting characteristics are
intimately related to topological and dimensional peculiarities
carried out by 3d space-time, like as the possibility of defining the
so-called Chern-Simons term, which provides remarkable novelties
whenever suitably added to a given model. For instance, the massless
degree of freedom (d.f.) which comes about from the pure Maxwell
action acquires a mass gap, without breaking the gauge symmetry,
when such an action is supplemented by the Chern-Simons term, leading
us to the so-called Maxwell-Chern-Simons (MCS) model, whose
Lagrangian reads:
\ba
{\cal L}_{MCS}=-\frac14 F^{\mu\nu}F_{\mu\nu}
+m\epsilon^{\mu\nu\kappa}A_\mu \partial_\nu A_\kappa \label{LMCS}\,.
\ea
As it is well-known, this mass generation mechanism
holds in Abelian and non-Abelian frameworks, and a similar one also
takes place in planar gravity (see Ref.\cite{Pgrav} for further
details).\\

Another interesting characteristic that takes place in
electrodynamic-like
models in (2+1) dimensions is the violation of the so-called Huyghens
principle by the classical (retarded and advanced) Green
functions associated to the free massless
vector field\footnote{Here, for the violation of Huyghens principle by such
functions, we are meaning that their mathematical support no longer lie only on
the surface of the light-cone, where $(x-y)^2=0$, but also spread
throughout other space-time intervals.}
(for details, see Refs.\cite{teseetc,CH,Baker,JJ}). This, in turn, was
shown to
lead us to some remarkable phenomena, such as the classical
reverberation of
electromagnetic signals and to a still missing Larmor-like
formula
relating the acceleration of the sources to the radiated power. Indeed,
these
aspects were verified in both Maxwell and
Maxwell-Chern-Simons electrodynamics (see Ref.\cite{teseetc}; see also 
Ref.\cite{Galtsov}).\\

In this work, we wish to go further into such a subject and deal
with
these points at the quantum level.
More precisely, we shall show that a similar violation takes also place
here.
This is explicitly demonstrated by showing that the Pauli-Jordan-like
commutators between two components of the field-strength has
non-vanishing support for light-like and time-like intervals as well,
i.e.,
$(x-y)^2\ge0$. However, although Huyghens principle no longer holds,
microcausality is kept, since these commutators are shown to vanish for
{\em every space-like interval}. Namely, such a result state us that the
measure of
the interference (correlation, more precisely) between the electric and
magnetic field will be non-vanishing whenever these fields be separated
by
non-space-like vectors.\\

Here, it is worthy noticing that such a result is in deep contrast
with (3+1)-dimensional facts, where such a correlation vanishes
unless the interval is light-like. The latter, as it is well-known, is
a
equivalent statement that (3+1)-dimensional quantized radiation
(genuine
photons) travels at speed $c$, what implies that they are
massless(for details, see Refs.\cite{Heitler,Bjorken,Sakurai,Jauch}). Actually, 
it is widely-known that, other physical
phenomena are deeply connected to these characteristics of the
radiation, like
as those processes concerning absorption, emission and scattering of
light.\\

Therefore, we may wonder whether some well-established
(3+1)-dimensional
results are still maintained unalterated in the planar case. As we have
already said, the ``matching'' between the validity of the Huyghens
principle
and the massless character of the radiation is a first result that must
be
abandoned, at least when (2+1)-dimensional facts are concerned. Indeed,
as we
shall see later, there are some emission and absorption processes that
have
to be also substancially modified whenever dealing with planar
electrodynamic-like
models. For instance, Planck law appears to
be proportional to the square frequency of the radiation,
$\nu^2$, instead of $\nu^3$ (the case in (3+1) dimensions).\\
\section{The Pauli-Jordan commutators and the violation of the Huyghens
principle} Although the Hamiltonian and equal-time commutation
relations constitute in the whole machinary necessary to
canonically
quantize the electromagnetic radiation and extract its physical
content,
sometimes it is very important calculate and explore the consequences
of
non-equal-time commutation relations between fields and/or conjugated
momenta.\\

For instance, we know that (3+1)-dimensional quantized radiation
(photons) has
vanishing CR's between two components of its potential unless the
space-time
interval is light-like, say:\footnote{We use the conventions:
$\hat{\mu},\hat{\nu}, {\mbox{\rm etc}}=0,1,2,3$ with
${\mbox {\rm diag}}(\eta_{\hat{\mu}\hat{\nu}})=(+,-,-,-)$. We also use
$\mu,\nu,{\mbox {\rm etc}}=0,1,2,$
while $i,j, {\mbox {\rm etc}}= 1,2 $. In addition,
${\mbox {\rm diag}}(\eta_{\mu\nu})=(+,-,-)$ and
$\epsilon^{012}=\epsilon_{012}= \epsilon^{12}=\epsilon_{12}=1$. We also denote 
by $\square$ the (2+1)-dimensional D'Alembertian differential operator: 
$\square=\partial_\mu\partial^\mu=\partial^2_t -\partial^2_i$.} 
\ba 
[A_{\hat{\mu}}(x),A_{\hat{\nu}}(y)]_{x^0\neq y^0}=-i \eta_{\hat{\mu}\hat{\nu}} 
D^{3+1}_{PJ}(x-y)\,,\label{crPJ4} \ea
where the massless Pauli-Jordan function is given by ($\Theta(z)$ is
the step-function):
\ba
D^{3+1}_{PJ}(z)=\frac{1}{4\pi}(\Theta(z^0)-\Theta(-z^0))\delta^{3+1}(z^2)
\,.\label{DPJ3+1}
\ea
Here, since the support of the function above lies only on the surface
of the
(3+1)-dimensional light-cone, we say that $D^{3+1}_{PJ}(x-y)$ satisfies
Huyghens' principle. In addition, it is not difficult to show that
relation (\ref{crPJ4}) implies, for example,
that ($\vE=-\nabla A^0-\partial_t\vA$ and $\vec{B}=\nabla\wedge\vA$,
as usual): \ba
[\vec{B}_1(x),\vec{E}_2(y)]_{x^0\neq
y^0}=i\frac{\partial^2}{\partial{x}^0\partial{y}^3}\,D^{3+1}_{PJ}(x-y)\,.
\label{crEB}
\ea
which is clearly non-vanishing if and only if $(x-y)^2=0$ (see
Refs.\cite{Heitler,Bjorken,Sakurai,Jauch}). Therefore, we conclude
that there is {\em interference} between measurements of
field-strength components if and
only if they are connected by light-like intervals. This statement, in
turn, is a direct consequence
from the massless character of the radiation, which implies that the
physical
excitations will always travel with velocity $c$, or in other words,
their
information may only be connected by light-like vectors. Therefore, the
statement of the massless feature of the Maxwell radiation is
equivalent to
state the validity of the Huyghens principle, by Pauli-Jordan function,
in
(3+1) dimensions.\\
\subsection{The massless case}As we shall see below, such a
``{\em matching}'' above will no longer hold
in (2+1) dimensions. More precisely, even though free Maxwell
Lagrangian (as usual, $F_{\mu\nu}=\partial_\mu A_\nu -\partial_\nu
A_\mu$), \ba
{\cal L}_M=-\frac14 F_{\mu\nu}F^{\mu\nu} \label{LMaxwell}\,,
\ea
describes a massless degree of freedom, $\square A_\mu=0$, we shall
show that a similar commutator between two components of the gauge
field (or of its associated field-strength) will be non-vanishing not
only if they were separated by light-like vectors, but also by time-like
ones, say, $(x-y)^2\ge 0$.\\

In order to see this, let us consider the
(2+1)-dimensional analogue of relation (\ref{crPJ4}), say:
\ba
& & [A_\mu(x),A_\nu(y)]_{x^0\neq y^0}=-i\eta_{\mu\nu}\,
D^{2+1}_{PJ}(x-y)\label{PJrel}\,.
\ea
Now, if we take the general form of the massless Pauli-Jordan
function,
\ba
D^{(d+1)}_{PJ}(z)=\frac{i}{(2\pi)^{d+1}}\int d^{d+1} k\,
\epsilon(k^0)\delta^{d+1}(k^2)\, \frac{e^{-ik_\mu z^\mu}}{k^2}\,,\label{PJqqdim}
\ea
with $\epsilon(k^0)=\Theta(k^0)-\Theta(-k^0)$ and
$\delta^{d+1}(k^2)=\delta^{d+1}(k_0^2 -\vk^2)$, and particularize it
to the planar case, we obtain:
\ba
D^{(2+1)}_{PJ}(z)=
\frac{1}{2\pi}[\Theta(z^0)-\Theta(-z^0)]\frac{\Theta(z^2)}{\sqrt{z_0^2-\vec
{ z}^2}} \,.\label{DPJ2+1}
\ea
Here, it is worthy noticing a remarkable feature of the function above:
as previously announced, it
yields a non-vanishing result whenever the space-time interval is
light-like or time-like, $z^2=(x-y)^2\ge0$. This is in deep contrast
with
$D^{3+1}_{PJ}$, compare with eq. (\ref{DPJ3+1}), whose support lies only
over light-like
vectors. Thus, we say that $D^{2+1}_{PJ}$ does not satisfy Huyghens
principle. Notice, however, that micro-causality is not lost:
$D^{2+1}_{PJ}$
vanishes whenever $z^2=(x-y)^2<0$. However, we {\em loose} the
``matching'' between the
statements of the massless character of a field and the validity of
Huyghens
principle as presented by its associated Pauli-Jordan
function in (2+1) dimensions.\\

Now, let us return to the (2+1)-dimensional Pauli-Jordan commutator.
From
expression (\ref{PJrel}) we may obtain similar relations between any
two
components of the field-strength. After some algebra we can show the
following
non-equal-time equality ($\tilde{F}^\mu=
\epsilon^{\mu\nu\kappa}\partial_\nu A_\kappa=(-B, -\epsilon^{ij}E^j)$,
as usual): \ba
[\tilde{F}^\mu (x), \tilde{F}^\nu(y)]=
-2i\left(\square\eta^{\mu\nu}
- \partial^\mu\partial^\nu\right)\,D^{(2+1)}_{PJ}(x-y)\label{PJF}\,,
\ea
which gives us, among other, that:
\ba
[E^i(x),B(y)]_{x^0\neq{y}^0}=+2i\epsilon^{ij}\frac{\partial^2}
{\partial y_j\partial x_0}\,D^{(2+1)}_{PJ}(x-y)=+3i
\frac{(x^0-y^0)\epsilon^{ij}(x^j-y^j)}{(\sqrt{(x-y)^2})^4}
\,D^{(2+1)}_{PJ}(x-y)\label{EB2+1}\,. \ea Clearly, the result above
states us that the {\em interference} between the
measurements of $\vec{E}$ and $B$ are non-vanishing whenever such
quantities
are separated by non-space-like intervals, $(x-y)^2\ge 0$.
\subsection{The topologically massive case}
In a standard massive theory, the non-equal-time CR's between two components
of the gauge potential would be read as:
\ba
[\phi_I(x),\phi_J(y)]_{x^0\neq y^0}=-ig_{IJ}{\cal D}^{d+1}(x-y;\,\mu)
\label{fifi}\,,
 \ea
where ${\cal D}^{d+1}$ is the (d+1)-dimensional massive Pauli-Jordan
function,
\ba
{\cal D}^{d+1}(z)=\frac{i}{(2\pi)^{d+1}}\int d^{d+1}k\,
\epsilon(k^0)\,\delta^{d+1}(k^2-\mu^2)\,\frac{e^{-ik^\mu 
z_\mu}}{(k^2-\mu^2)}\label{calD}\,, \ea
which smoothly recover its massless counterpart as $\mu\to0$.\\

Nevertheless, whenever topological-like mass generation mechanisms are
involved, then relation (\ref{fifi}) may acquire new terms. This is the
case for MCS model, eq. (\ref{LMCS}). Actually, due to the
Chern-Simons term, a similar expression to (\ref{fifi}), for
$A^\mu$-potential, gets the following form (see, for example,
Ref.\cite{japas} for further details): \ba
&[A^\mu(x),A^\nu(y)]_{x^0\neq
y^0}=&-i\left(\eta^{\mu\nu}+\frac{\partial^\mu
_x\partial^\nu_y}{m^2}-\frac{\epsilon^{\mu\nu\kappa}\partial_\kappa}{m}
\right){\cal D}^{2+1}(x-y;m)\,+\nn\\  & & +\frac{i}{m^2}
\left(\partial^\mu_x\partial^\nu_y
-m\epsilon^{\mu\nu\kappa}\partial_\kappa\right)
\,D^{2+1}_{PJ}(x-y)\,,\label{AAMCS} \ea whose first term is the
contribution due to the massive physical excitation, while the second
one comes from the (non-dynamical) massless pole. The latter answers for the
magnetic vortex-like flux attached to the electric charge. Notice also
that as $m\to0$ relation (\ref{AAMCS}) reduces to its massless
counterpart, eq.(\ref{PJrel}).\\

Now, taking eq. (\ref{calD}) for the (2+1)-dimensional case, we
obtain that:
\ba
{\cal D}^{2+1}(z)=\frac{1}{2\pi}
[\Theta(z^0)-\Theta(-z^0)]\,\frac{\cos(m\sqrt{z^2 })\,
\Theta(z^2)}{\sqrt{z^2}} \,,\label{calD21}
\ea
which is clearly
non-vanishing for all $z^2\ge 0$. Hence, we readily see that the function 
above does not satisfy Huyghens principle (but respects
micro-causality, since it identically vanishes for all $z^2<0$).
However, such a `{\em violation}' is not so drastic here. It was
indeed expected to happen, since as we know very well, massive
excitations are hadrons, thus propagating slower than light. Indeed,
similar `{\em violation}' may also take place in (3+1) dimensions
whenever dealing with massive particles (see, for example,
Ref.\cite{Barut}).\\

Although both massless and massive poles contribute to expression
(\ref{AAMCS}), when we calculate a similar relation for the field-strength
components, we are led to (notice that, as $m\to0$ the commutator below recovers 
its massless counterpart, (\ref{PJF})) \ba
[\tilde{F}^\mu(x),\tilde{F}^\nu(y)]_{x^0\neq
y^0}=i\,(\square\eta^{\mu\nu}-\
\partial^\mu\partial^\nu\,+m\epsilon^{\mu\nu\kappa}\partial_\kappa)\,
{\cal D}^{2+1}(x-y;m)\,,\label{FFMCS}
\ea
which highlights that only the massive pole gives us a non-vanishing
value when the electric and magnetic fields are measured. In other words,
only the massive degree of freedom can be properly detected as a physical
excitation (see also Ref.\cite{japas}). It is also clear that, even though
in a quite particular way, the {\em interference} between measurements of
the field-strength will give a non-trivial result whenever the space-time
interval that separates them is light-like or time-like, say, $(x-y)^2\ge
0$. For example:
\ba
& [E^i(x),B(y)]_{x^0\neq y^0}=& i\left[\frac{(x_0-y_0)
\epsilon^{ij}(x^j-y^j)}{(\sqrt{(x-y)^2})^2}\left(m^2
+3\frac{\tan(m\sqrt{(x-y)^2})}{\sqrt{(x-y)^2}}+\frac{3}{(\sqrt{(x-y)^2
} )^2} \right) \right .\nn \\ & & \left.
+m\frac{(x^i-y^i)}{\sqrt{(x-y)^2}}\left(m\tan(m\sqrt{(x-y)^2})+
\frac{1}{\sqrt{(x-y)^2}}\right)\right]{\cal D}^{2+1} (x-y,m)\nn
\ea
which, despite its rather complicated form is clearly
non-vanishing whenever $(x-y)^2\ge0$, what confirms the fact that
massive radiation is detected as propagating slower than light. In
addition, it is worthy noticing that the last term in expression
above, proportional to $m(x^j-y^j)$, comes from the Chern-Simons
term and thus is odd under spin flipping, say, $m\to-m$. Notice also
that as $m\to 0$, expression above recover its massless counterpart,
eq. (\ref{EB2+1}).\\
\section{Emission and absorption of planar photons}Now, we shall study
some physical processes concerning emission and absorption of planar
radiation by non-relativistic atomic electrons, namely, how
Einstein-like coeficients and Planck read in the planar world. Notice, 
hereafter, the explicit presence of the constants $h$ and $c$.\\

First, let us consider an atom in a initial state
$A$, which may interact with a $n$-photon state,
$n_{\vec{k},\vec{\xi}}$ with momentum
$\vec{k}$ and polarization vector $\vec{\xi}$. Now, the transition
matrix element for absorption of an unique photon (first-order
process) reads:
\ba
<B;n_{\vec{k},\vec{\xi}}-1|H^{(1)}_{int}|A;n_{\vec{k},\vec{\xi}}>=-
\frac{e}
{m}\sqrt{\frac{n_{\vec{k},\vec{\xi}}}{2\omega(\vec{k})}}\,<B|\sum_i
e^{i\vec{k}\cdot\vx_i}\vec{\xi}(\vec{k})\cdot\vec{p}_i|A>\,,
\ea
which similarly to its (3+1)-dimensional analogue, is proportinal to
the squared root of the total number of photons. In expression above,
$H^{(1)}_{int}$ is the (1st order) interaction Hamiltonian between
photons and electrons, given by:$$H^{(1)}_{int}=-\frac{e}{m}\sum_i
\vA(\vx_i,t)\cdot\vec{p}_i,$$ with $\vec{p}_i$ being the momentum of
the i-th electron located at $\vx_i$, where $\vA(\vx_i,t)$ acts at
time $t$. The field $\vA$, in turn, is decomposed in plane-waves:
\ba
A^\mu(\vx,t)=\frac{1}{(2\pi)^2}\int
\frac{d^2\vk}{\sqrt{2\omega(\vk)}}\xi^\mu(\vk)\,[\bfa(\vk)e^{-ik^\nu
x_\nu}
+\bfa^\dagger(\vk)e^{+ik^\nu x_\nu}] \label{pwavesless}\,,
\ea
with $\xi^\mu(\vk)$ being the polarization vector, whose explicit
form reads like below:
\ba
\xi^\mu(\vec{k})=(0;1,0)\label{ximassless}\,,
\ea
whenever the radiation is massless, or else (at rest frame -Landau
gauge) \ba
\xi^\mu_L(\vec{k}=0)=\left(0;\,
\frac{1}{\sqrt{2}}\,,\,-\frac{i}{\sqrt{2}}\frac{m}{|m|}\right)\label{zeta12}\,,
\ea
for photons appearing in the MCS framework. In the latter case, such a
structure of the polarization vector has provided a natural (say,
Lorentz and gauge covariant) explanation why such excitations display
spin $\pm 1$ (see
Refs.\cite{DeVechi,Banerjee}, for further details). \\

For first-order emission processes, we also have:
\ba
<B;n_{\vec{k},\vec{\xi}}+1|H^{(1)}_{int}|A;n_{\vec{k},\vec{\xi}}>=-\frac{e}
{m}\sqrt{\frac{n_{\vec{k},\vec{\xi}}+1}{2\omega(\vec{k})}}\,<B|\sum_i
e^{i\vec{k}\cdot\vx_i}\vec{\xi}(\vec{k})\cdot\vec{p}_i|A>\,.
\ea
It is worth noticing that result above coincide, in form, with usual
(3+1)-dimensional ones, for both cases, {\em stimullated} ($n\neq0$)
and {\em spontaneous} ($n=0$) emissions. Namely, for $n=0$ we
realize that the spontaneous emission of planar photons takes place,
like in the 3-spatial case: the transition matrix element is
non-vanishing although no radiation is present.\\

Now, we may readily show that, while matrix elements differ whenever the
photon carries mass or not, since they strongly depend on the structure of
the polarization vector, eqs. (\ref{ximassless}) and (\ref{zeta12}),
the probability transition, basicaly
$|<B;n\pm1|H^{(1)}_{int}|A;n>|^2$, may be shown to give the same
results, independing on the massive character. At this point we
should then stress that such a quantity is not modified by the
(topological) mass gap of the excitations, at least in the planar
case. This is because such a gap does not increase the number of
degrees of freedom of the planar radiation.\\

Going further, and searching for how Planck law reads in the planar
case, we find that:
\ba
U^{2+1}(\nu)=2\pi\frac{h\nu^2}{c^2}\frac{1}{(e^{h\nu/KT}-1)}
\,,\label{U2+1} \ea which contrasts with its (3+1)-dimensional
counterpart, \ba
U^{3+1}(\nu)=8\pi\frac{h\nu^3}{c^3}\frac{1}{(e^{h\nu/KT}-1)}
\,,\label{U3+1} \ea
by a multiplicative factor of $4\,\nu/c$. Indeed, it is not difficult
show that such a factor arises by virtue of dimensional matters, like as
the ``volume element'' and the number of physical degrees of freedom
carried by massless radiation in both space-time. As it is
well-known, $U(\nu)$ represents the energy density\footnote{Essentially, 
$U(\nu)$ is the total energy enclosed in a given ``volume'' times the density of
allowed physical states (per unity frequency),
${\rho}_{\nu,d\nu}=N\cdot{\prod}_{i} dk_i$, $\vec{k}$
is the wave-vector and $N$ the number of degrees of freedom carried by
the excitations.}  (per unity
frequency, $d\nu$) of radiation in thermal equilibrium, at temperature
$T$, distributed over a given ``volume'', with frequency ranging from
$\nu$ to $\nu+d\nu$ (see Refs. \cite{Heitler,Bjorken, Sakurai} for further 
details). Moreover, comparing eqs. (\ref{U2+1}) and (\ref{U3+1}), we easily 
realize that ultra-violet divergences concerning electromagnetic radiation is 
better handled in planar world than in 3D-spatial case.\\

Furthermore, working on the topologically massive version of the
planar Planck law, we must recall that the excitations now present
rest energy, say, $E^2=|\vec{p}|^2c^2+m^2c^4,$ what implies that the
frequency of the massive photons, $\nu'=\omega'/2\pi=E/h$ is no longer
determined only by its wave-vector, $\vec{k}=\vec{p}/h$, but also by
its mass. Actually, following the same steps like in the previous
case, we find
\ba
U^{2+1}(\nu',m)=2\pi\frac{h{\nu'}^2}{c^2}\frac{1}{(e^{h\nu'/KT}-1)}
\,,\label{U2+1m}
\ea
with
$\nu'=c\sqrt{(\vec{k}/2\pi)^2+(mc/\hbar)^2}=c\sqrt{\nu^2+\nu^2_0}$,
where $\nu$ and $\nu_0$ are the kinetic and ``rest'' frequencies of
the massive radiation. Then, we now understand $U^{2+1}(\nu,m)$ as
the distributed energy density (per unity frequency, $d\nu'$) of
photons, in thermal equilibrium, with frequency between $\nu'$ and
$\nu'+d\nu'$, and (constant) physical mass $m$.\\

In addition, we may easily show that in the small mass limit, $|m|<<1$
or equivalently, $\nu>>\nu_0$, eq. (\ref{U2+1m}) gives
us$$U^{2+1}(\nu,m)_{|m|<<1}=2\pi\frac{h{\nu}}{c^2}\frac{\nu+\nu_0}
{(e^{h\nu/KT}-1)}+{\cal O}(\nu^2_0)\,,$$which recovers the massless
result, eq. (\ref{U2+1}), as $m\to0$.
\section{Conclusions and Prospects} We have shown
that Huyghens principle is not satisfied by Pauli-Jordan-like
functions in (2+1) dimensions (massless or massive cases). In the
massless scenario, this has led us to the lost of the matching which
connects the massless character of photons and its propagation
at light-speed, $c$. Indeed, by applying to the quantum framework, a
similar interpretation to that we have done in the classical case
(see Ref.\cite{teseetc}), we are led to some surprisingly questions,
for instance, the following ones. First, by facing an electromagnetic
signal rather as a wave, reverberation affects its propagation and we
can no longer speak of sharp pulses. On the other hand, whenever
quantizing such a wave in order to give it the status of a particle (a
``planar photon''), we may wonder whether the concept of a photon as a
localized energy packet should not be reassessed in the planar case.
Second, since $A_\mu$ describes an elementary particle (an irreducible
representation of the $SO(2,1)$ Lorentz-like group), how could we
conciliate such a structureless character with the division of the
planar massless photons into minor parts?\\

We have also seen that probability transition elements for
emission and absorption of planar photons by non-relativistic atomic electrons
do not generally change, neither by lowering the dimension of space-time,
(3+1)D to (2+1)D, nor by virtue of the (topological) mass gap. On the
other hand, we have seen that Planck law has to be substancially
modified in the planar world. For instance, in the massless case a
frequency-dependent multiplicative factor was found to contrast it
from its 4-dimensional counterpart. Moreover, the topologically
massive case was worked out and shown to recover massless result as
the mass parameter vanishes.\\

At this point, we should also discuss about possible conditions that
could provide the dynamics of genuine photons as they were planar
ones. Indeed, under certain circumstances (low temperature, high
magnetic field, and so forth) electrons are shown to perform a
(quasi) planar dynamics, which is the case when quantum Hall effect
and high-Tc superconductivity come about. Therefore, we may wonder
whether some physical conditions, unkown to us at the present, could
lead us to a similar scenario for photons. For instance, inside a
superconductor vector bosons acquire mass by virtue of the Meissner
effect. Now, if we intend to describe the electrodynamics inside such 
a sample by means of Maxwell-Chern-Simons model (the anyonic
superconductivity proposal, for example), then the observed vector
gauge excitations should be expected to behave, at some extent, as
much as they were ``planar photons''. Perhaps, some experiments
towards the measurement of the interference between electric and
magnetic fields associated to such excitations could yield to results
well-fit by massive Pauli-Jordan commutators; or still they could
determine the energy distribution of a ``bath'' of photons and confirm
the present results concerning this isssue.\\

\begin{sloppypar}In forthcoming communications, we intend to go further into the
abovementioned analysis and investigate possible modifications on
other physical results whenever dealing with (2+1)d
electrodynamic-like models, namely those subjects concerning emission,
absorption, and scattering of radiation.\\ \\ \end{sloppypar}

\centerline{\bf \Large Acknowledgements}
W.A.M-M thanks CBPF and GFT/UCP, where part of this work was done. He is also
grateful to Funda\c{c}\~ao de Amparo \`a Pesquisa do Estado de Minas Gerais (FAPEMIG)
and Conselho Nacional de Desenvolvimento Cient\'{\i}fico e Tecnol\'{o}gico (CNPq)
for the financial support. J.A.H-N thanks CNPq for partial financial support.


\begin{thebibliography}{99}
\bibitem{DJTetc}W. Siegel, Nucl.Phys. {\bf B156} (1979) 135; J. Schonfeld,
{\em ibid} {\bf B185} (1981) 157; S. Deser, R. Jackiw and S. Templeton,
Ann. Phys. {\bf 140} (1982)372; Phys. Rev. Lett. {\bf 48} (1982) 975. See
also, R. Jackiw' Lectures Notes V$^{\mbox{\rm th}}$ J.A. Swieca Summer
School '89 (Sociedade Brasileira de F\'{\i}sica); R. Jackiw, {\em
Topics in Planar Physics} [unpublished].
 \bibitem{qhe}R. Prange and S. Girvin, {\em The Quantum Hall Effect}
(Springer, New York, 1987); H. Aoki, Rep. Progr. Phys. {\bf 50} (1987)
655; G. Morandi, {\em Quantum Hall Effect} (Bibliopolis, Naples, 1988).
\bibitem{htcsup}R. Laughlin, Phys. Rev. Lett. {\bf 60} (1988) 2617; A.
Fetter, C. Hanna and R. Laughlin, Phys. Rev. {\bf B39} (1989) 9679; S.
Randjbar, A. Salam and J, Strathdee, Int. J. Mod. Phys. {\bf B5} (1991)
845; N. Dorey and N. Mavromatos, Phys. Lett. {\bf B266} (1991) 163.
\bibitem{anyonsup}F. Wilczek, {\em Fractional Statistics and Anyon
Superconductivity} (World Scientific, Singapore, 1990).
\bibitem{frac}F. Wilczek, Phys. Rev. Lett. {\bf 48} (1982) 1144. See
also A. Lerda, {\em Anyons: Quantum Mechanics of Particles with
fractional Statistics}, Lectures Notes in Physics (Springer
International, Berlin, 1992). \bibitem{Pgrav}See S. Deser, R. Jackiw
and S. Templeton in Ref.\cite{DJTetc}.
\bibitem{teseetc}W.A. Moura-Melo and J.A. Helay\"el-Neto, Phys. Rev. {\bf
D63} (2001) 065013; W.A. Moura-Melo, Ph.D. Thesis, CBPF, 2001.
\bibitem{CH}R. Courant and D. Hilbert, {\em Methods of Mathematical
Physics} (Interscience Publishers, NY, 1962). In Vol. 2, see mainly
pp. 202-206 and 760-766.
\bibitem{Baker}B. Baker and E. Copson, {\em The Mathematical Theory of
Huyghens Principle} (Oxford Univ. Press, London, 1953).
 \bibitem{JJ}  J.J. Giambiagi, Il Nuovo Cimento {\bf A104} (1991)1841; C.G.
Bollini and J.J. Giambiagi, J. Math. Phys. {\bf 34} (1993)610.
\bibitem{Galtsov}D.V. Gal'tsov, {\em Radiation Reaction in Various Dimensions}, 
hep-th/0112110.
\bibitem{Heitler}W. Heitler, {\em The Quantum Theory of 
Radiation} (Oxford Univ. Press, 1944).
\bibitem{Bjorken}J.D. Bjorken and S. Drell, {\em Relativistic Quantum
Fields} (McGraw-Hill, NY, 1965)
\bibitem{Sakurai}J.J. Sakurai, {\em Advanced Quantum Mechanics}
(Addison-Wesley, Reading, 1967).
\bibitem{Jauch}J. Jauch and F. R\"orhlich, {\em The Theory of Photons and
Electrons} (Addison-Wesley, Reading, 1959).
\bibitem{DeVechi}F. Devecchi, M. Fleck, H. Girotti, M. Gomes and A. Da
Silva, Ann. Phys. {\bf 242} (1995) 275.
\bibitem{Banerjee}R. Banerjee, B. Chakraborty and T. Scaria, Int. J. Mod.
Phys. {\bf A16} (2001) 3967.
\bibitem{japas}T. Kimura, Prog. Theor. Phys. {\bf 81} (1989) 1109; N. Imai,
K. Ishikawa and I. Tanaka, {\em ibid} {\bf 81} (1989) 759; N. Nakanishi,
Int. J. Mod. Phys. {\bf A4} (1989) 1055.
\bibitem{Barut}A.O. Barut, {\em Electrodynamics and Classical Theory of
Fields and Particles} (Dover Edition, NY, 1980). See mainly pp. 158-160.
\end{thebibliography}
\end{document}